# Robust and Sensitive Method of Lyapunov Exponent for Heart Rate Variability

Mazhar B. Tayel[1] and Eslam I AlSaba[2]

[1,2] Department of Electrical Engineering, Alexandria University, Alexandria, Egypt
profbasyouni@gmail.com
eslamibrahim@myway.com

## ABSTRACT

*Heart Rate Variability (HRV) plays an important role for reporting several cardiological and non-cardiological diseases. Also, the HRV has a prognostic value and is therefore quite important in modelling the cardiac risk. The nature of the HRV is chaotic, stochastic and it remains highly controversial. Because the HRV has utmost importance, it needs a sensitive tool to analyze the variability. In previous work, Rosenstein and Wolf had used the Lyapunov exponent as a quantitative measure for HRV detection sensitivity. However, the two methods diverge in determining the HRV sensitivity. This paper introduces a modification to both the Rosenstein and Wolf methods to overcome their drawbacks. The introduced Mazhar-Eslam algorithm increases the sensitivity to HRV detection with better accuracy.*

## KEYWORDS

*Heart Rate Variability, Chaotic system, Lyapunov exponent, Transform domain, and Largest Lyapunov Exponent.*

## 1. INTRODUCTION

Cardiovascular diseases are a growing problem in today's society. The World Health Organization (WHO) reported that these diseases make up about 30% of total global deaths. Those heart diseases have no geographic, gender or socioeconomic boundaries [3]. Therefore, early stage detection of cardiac irregularities and correct treatment are very important. This requires a good physiological understanding of cardiovascular system.

Studying the fluctuations of heart beat intervals over time reveals a lot of information called heart rate variability (HRV) analysis. A reduction of HRV has been reported in several cardiological and non-cardiological diseases. In addition, HRV has a prognostic value and is therefore quite significant in modelling the cardiac risk. HRV has already proved his usefulness and is based on several articles that have reviewed the possibilities of HRV [1 - 5].

The fact that HRV is a result of both linear and nonlinear fluctuations opened new perspectives as previous research was mostly restricted to linear techniques. Some situations or interventions can change the linear content of the variability, while leaving the nonlinear fluctuations intact. In addition, the reverse can happen: interventions, which up till now have been believed to leave cardiovascular fluctuations intact based on observations with linear methods, can just as well modify the nonlinear fluctuations. This can be important in the development of new drugs or treatments for patients. This paper introduces a modification algorithm method to overcome the drawbacks arising in both Rosenstein and Wolf methods using the same approach of Lyapunov exponent. That analysis the nonlinear behaviour of the HRV signals. The modified method create a new chapter of sensitivity of HRV by a new description for Lyapunov exponent. The modified Mazhar-Eslam method achieve more accuracy than another Lyapunov methods Wolf and Rosenstein.

## 2. HEART RATE VARIABILITY

Heart Rate Variability (HRV) is a phenomenon that describes temporal variation in intervals between consecutive heartbeats in sinus rhythm. HRV refers to variations in beat-to-beat intervals corresponding to instantaneous HRs. HRV is a reliable reflection of many physiological factors modulating normal rhythm of heart. In fact, they provide a powerful means of observing interplay between sympathetic and parasympathetic nervous systems. It shows that a structure generating a signal is not only simply linear, but also involves nonlinear contributions.

Spontaneous variability of HR has been related to three major physiological originating factors: quasi-oscillatory fluctuations thought to arise in blood-pressure control, variable frequency oscillations due to thermal regulation, and respiration. Frequency selective analysis of cardiac inter-beat interval sequences allows separate contributions to be isolated. Using this method, a laboratory and field study of effects of mental work load on the cardiac interval sequence has been carried out [4].

The diagnosticity of HR is restricted by several factors like environmental stressors and physical demands that may be associated with a task. These tasks may have different physiological consequences and change in HR may depends on these factors more than mental workload. Backs (1998) [6] focused on the fact that observed HR could be caused by different underlying patterns of autonomic nervous system activity. If different information processing demands affect the heart via different modes of autonomic control, it could increase diagnosticity of HR [6]. Backs" study addressed the validity of the autonomic component, using data from a large study, in which many central and peripheral psycho-8 physiological measures were collected simultaneously while performing single and dual tasks which had different physical demands. The measures collected were the residual HR, parasympathetic and sympathetic activity, respiratory sinus arrhythmia (RSA), and THM (Traube-Hering-Mayer) wave using principle component analysis (PCA), image factoring, impedance cardiogram (ZKG) and ElectroCardioGram (ECG). The sympathetic and parasympathetic systems were examined for independence. From the study, the PCA factors computed on raw ECG data provided useful information like different autonomic modes of control were found that were not evident in heart period. The objective was to verify if factors extracted using residual HR as a marker variable validly reflected cardiac sympathetic activity and if the solutions obtained from raw and baseline corrected data were in compliance with each other. This information about the underlying autonomic activity may increase the diagnosticity of HR. Figure 1 shows the example of HR variation of normal subject (control case).

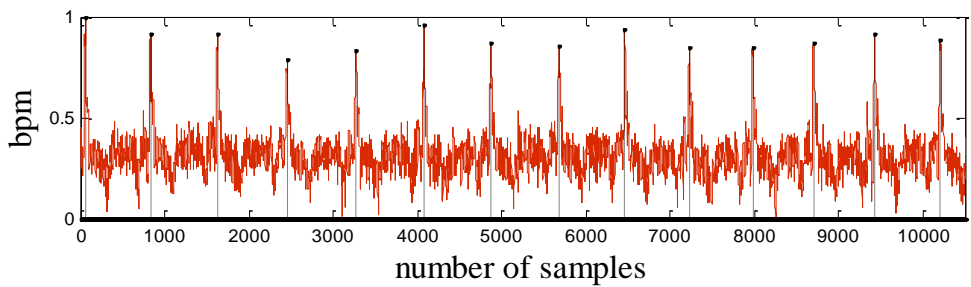

Figure 1. Heart rate variation of normal subject.

## 2.1 Components of Heart Rate

HRV was originally assessed by calculating the mean beat to beat heart rate commonly called the RR interval, and its standard deviation measured on short term ECG. To date about 26 different types of arithmetic manipulations of RR intervals have been used . Variable of interest for the study was the RMSSD index. The RMSSD index is defined as the root mean square of the differences of the successive RR intervals. MAX-MIN or peak-valley quantification of HRV is the difference between the shortest RR interval during inspiration and longest during expiration. RMSSD is a time domain measure of HRV. Recently spectral analysis of HRV has been developed which transforms the signal from time domain to frequency domain. The power spectrum of heart rate and blood pressure yields three major bands. A low frequency peak ranging between .06 Hz to .15 Hz, a high frequency peak rangingbetween .15 Hz to .4 Hz and a very low frequency peak below .05 Hz constitute the power spectrum [13]. The LF is associated with blood pressure control, reflecting sympathetic activity. The HF is correlated with respiratory sinus arrhythmia reflecting parasympathetic activity [13]. The VLF is linked with vasomotor control or temperature control. The RR interval data obtained from ECG or other heart rate monitoring devices can be analyzed using any mathematical tool like Fourier transformation or moving average method [14]. Statistical significance can be tested using standard tests after the frequency domain analysis. For the current study, in the first phase the raw data obtained from the heart rate monitor was analyzed using both the time domain and the frequency domain measures. The data further obtained from these analyses was tested for statistical significance using analysis of variance (ANOVA).

## 2.2 Heart Rate Measures

Important information on the heart rate components can be summarized from the as the components of HRV are markers of the sympathetic and parasympathetic activity. The low frequency component is associated with blood pressure control or sympathetic activity. In addition, it is high, when the person is in high strain conditions. The high frequency component is an indicator of parasympathetic autonomic response. Also, it is an indicator of respiratory sinus arrhythmia (RSA). Besides, it is reduced during heavy exertions and awkward postures. The ratio of low to high frequency is an estimate of mental stress. It is used as an index of parasympathetic and sympathetic balance. This ratio is correlated with high job strain, when high. The very low frequency is linked with temperature control. It is Often seen as an unreliable measure.

Reduced HRV predicts sudden death and is a marker of fatal ventricular arrhythmia. This reduction in HRV can be examined from time domain components like the RMSSD. All of these components are responsible for the identification of increased cardiac reactivity in individuals under high stress.

## 2.3 Practical Applications of Heart Rate Monitors

HRV is sensitive to both physiological and psychophysical disorders. In recent years HRV has also been used as a tool to improve diagnosing of heart rate in the general population which included both the working and non-working population. Assessing the impact of physical and mental demands associated with tasks in a work place on the heart can help to predict cardiac diseases. Therefore it has become essential to measure HRV [15].

Measurement of HRV in the past required a high-quality electrocardiogram (ECG), but the cost and complexity of the ECG equipment has made it difficult to perform HRV analysis particularly in the physical training field [16]. To address these needs, several portable heart rate monitoring devices have been designed. The development of wireless heart rate monitoring (HRM) with elastic electrode belt allowing the detection of RR intervals with a resolution of 1ms represents an interesting alternative to classic fixed or ambulatory ECG for use outside of laboratory settings.

Kingsley demonstrated that the R-R intervals and heart rate measurements obtained using Polar 810S and ambulatory ECG system are in good agreement with each other [16]. Gamelin examined the validity of the Polar S810 monitor to measure RR intervals at rest. Narrow limits of agreement, good correlation and small effect sizes validate the monitor and measure RR intervals to make HRV analysis [15]. Sandercock measured the reliability of three commercially available HRV instruments using short term recordings. These recordings were made in three conditions: lying supine, standing, and lying supine with controlled breathing. Reliability was calculated using CV (coefficient of variation), ICC (intraclass correlation coefficient) and LoA (limits of agreement). The study showed supine condition is more reliable. On the whole the short term recordings less than five minutes were seen to be unreliable [17].

On the whole heart rate monitors pave a simple pathway to measure the heart rate in a feasible manner as they are affordable, portable, reliable for recordings more than 5 minutes and also easy to handle. For this study the Polar RS800 monitor (Polar Electro Inc., Lake Success, NY) was used to record the data. It has a chest strap and a watch which transmits data wirelessly. Previous literature reviews showed that various heart rate monitors had been checked for their validity and reliability. But the comfort levels experienced while using the heart rate monitor had not been addressed. This is important because wearing a heart rate monitor in a real time situation during work should not affect the workers performance and should be as comfortable as possible. One of the purposes of this study was to see how comfortable the wearer feels while doing various tasks, so that the HR monitor can be used for real time recording. Also statistical consistency had not been tested thoroughly and there is little research on HRV during a physical or manual activity like moving something heavy. Earlier studies had shown that the heart rate increases when a manual material handling job is performed, but the spectral components had not been studied in association with predictability of diseases [18].

## 3. LYAPUNOV EXPONENT

Lyapunov exponent $\Lambda$ is a quantitative measure of the sensitive dependence on the initial conditions. It defines the average rate of divergence or convergence of two neighbouring trajectories in the state-space. Consider two points in a space, $X_0$ and $X_0 + \Delta x_0$, each of which will generate an orbit in that space using system of equations as shown in figure 2.

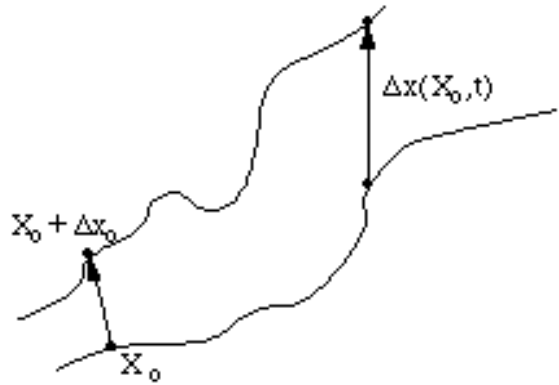

Figure 2. Calculation for two neighbouring trajectories.

These orbits are considered as parametric functions of a variable. If one of the orbits is used as a reference orbit, then the separation between the two orbits will be a function of time. Because sensitive dependence can arise only in some portions of a system, this separation is a location function of the initial value and has the form $\Delta x(X_0, t)$. In a system with attracting fixed points or attracting periodic points, $\Delta x(X_0, t)$ decreases asymptotically with time. For chaotic points, the function $\Delta x(X_0, t)$ will behave unpredictably. Thus, it should study the mean exponential rate of divergence of two initially closed orbits using the formula

$$\Lambda = \lim_{\substack{t \to \infty \\ |\Delta x_0| \to 0}} \frac{1}{t} \ln \frac{\Delta x(X_0, t)}{|\Delta x_0|} \qquad (1)$$

This number, called the Lyapunov exponent "$\Lambda$". An exponential divergence of initially nearby trajectories in state-space coupled with folding of trajectories, to ensure that the solutions will remain finite, is the general mechanism for generating deterministic randomness and unpredictability. Therefore, the existence of a positive $\Lambda$ for almost all initial conditions in a bounded dynamical system is the widely used definition of deterministic chaos. To discriminate between chaotic dynamics and periodic signals, $\Lambda$ s are often used. The trajectories of chaotic signals in state-space follow typical patterns. Closely spaced trajectories converge and diverge exponentially, relative to each other. *A negative exponent*$(\Lambda < 0)$ the orbit attracts to a stable fixed point or stable periodic orbit. Negative Lyapunov exponents are characteristic of dissipative or non-conservative systems. Such systems exhibit asymptotic stability. The more negative the exponent, the greater the stability. Super stable fixed points and super stable periodic points have a Lyapunov exponent of $\Lambda = -\infty$. This is something similar to a critically damped oscillator in that the system heads towards its equilibrium point as quickly as possible. A *zero exponent*$(\Lambda = 0)$ the orbit is a neutral fixed point (or an eventually fixed point). A Lyapunov exponent of zero indicates that the system is in some sort of steady state mode. A physical system with this exponent is conservative. Such systems exhibit Lyapunov stability. Take the case of two identical simple harmonic oscillators with different amplitudes. Because the frequency is independent of the amplitude, a phase portrait of the two oscillators would be a pair of concentric circles. The orbits in this situation would maintain a constant separation. Finally, a *positive exponent* implies the orbits are on a *chaotic attractor.*Nearby points, no matter how close, will diverge to any arbitrary separation. These points are unstable [3]. The flowchart of the practical algorithm for calculating largest Lyapunov exponentsis shown in figure 3.

In the following, the Largest Lyapunov Exponent algorithms are discussed.

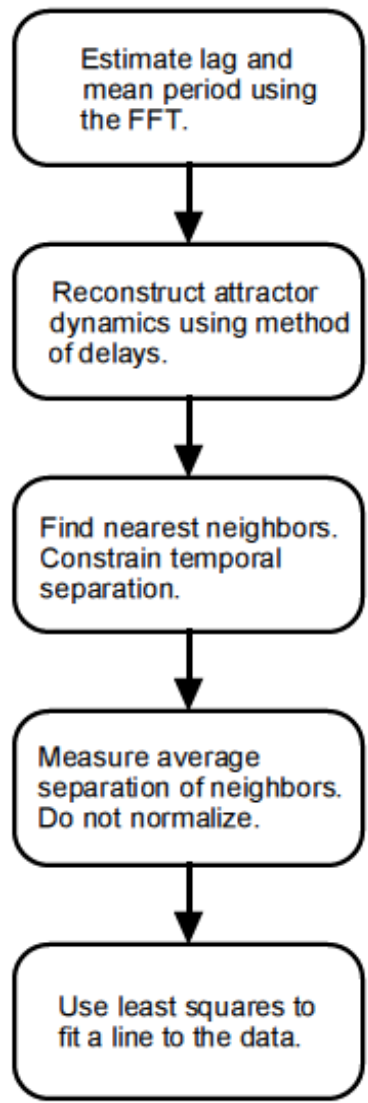

Figure 3. Flowchart of the practical algorithm for calculating largest Lyapunov exponents

### 3.1. Wolf's Algorithm

Wolf's algorithm [7] is straightforward and uses the formulas defining the system. It calculates two trajectories in the system, each initially separated by a very small interval. The first trajectory is taken as a reference, or 'fiducial' trajectory, while the second is considered 'perturbed'. Both are iterated together until their separation is large enough.

In this study, the two nearby points in a state-space $x_i$ and $x_i + \Delta x$, that are function of time and each of which will generate an orbit of its own in the state, the separation between the two orbits Δx will also be a function of time. This separation is also a function of the location of the initial value and has the form $\Delta x(x_i, t)$, where $t$ is the value of time steps forward in the trajectory. For chaotic data set, the mean exponential rate of divergence of two initially close orbits is characterized by

$$\Lambda = \lim_{K \to \infty} \frac{1}{K} \ln \frac{|\Delta x(x_i,t)|}{|\Delta x|} \tag{2}$$

The maximum positive Λ is chosen as $\Lambda_m$.

Largest Lyapunov exponent's quantify sensitivity of the system to initial conditions and gives a measure of predictability. This value decreases for slowly varying signals like congenital heart block (CHB) and Ischemic/dilated cardiomyopathyand will be higher for the other cases as the variation of RR is more [8, 9]. The LLE is 0.505 for the HR signalshown in Figure 1.

### 3.2. Rosenstein's Algorithm

Rosenstein's algorithm [10] works on recorded time-series, where the system formulas may not be available. It begins by reconstructing an approximation of the system dynamics by embedding the time-series in a phase space where each point is a vector of the previous m points in time (its 'embedding dimension'), each separated by a lag of j time units. Although Taken's theorem [11] states that an embedding dimension of 2D + 1 is required to guarantee to capture all the dynamics of a system of order D, it is often sufficient in practice to use m = D. Similarly, although an effective time lag must be determined experimentally, in most cases j = 1 will suffice.

Given this embedding of the time-series, for each point founding its nearest neighbour (in the Euclidean sense) whose temporal distance is greater than the mean period of the system, corresponding to the next approximate cycle in the system's attractor. This constraint positions the neighbours as a pair of slightly separated initial conditions for different trajectories. The mean period was calculated as the reciprocal of the mean frequency of the power spectrum of the time-series calculated in the usual manner using the Fast Fourier Transform (FFT). As shown in figure 3.

Now it can be performed a process similar to Wolf's algorithm to approximate the Largest Lyapunov Exponent (LLE). The first step of this approach involves reconstructing the attractor dynamics from the RR interval time series. After reconstructing the dynamics, the algorithm locates the nearest neighbour of each point on the trajectory. The nearest neighbour, $x_j'$, is found by searching for the point that minimizes the distance to the particular reference point, $x_j$. This is expressed as:

$$d_j(0) = \min ||x_j - x_{j'}|| \qquad \forall x_{j'} \tag{3}$$

where $d_j(0)$ is the initial distance from the $j^{th}$ point to its nearest neighbour and || ... || denotes the Euclidean norm. An additional constraint is imposed, namely that nearest neighbours have a temporal separation greater than the mean period of the RR interval time series. Therefore, one can consider each pair of neighbours as nearby initial conditions for different trajectories. The largest Lyapunov exponent (LLE) is then estimated as the mean rate of separation of the nearest neighbours. More concrete, it is assumed that the $j^{th}$ pair of nearest neighbours diverge approximately at a rate given by the largest Lyapunov exponent $\Lambda_m$:

$$d_j(j) \approx d_j(0) e^{\Lambda_m(i.\Delta t)} \tag{4}$$

By taking the ln of both sides of this equation:

$$\ln d_j(j) \approx \ln d_j(0) + \Lambda_m(i.\Delta t) \tag{5}$$

which represents a set of approximately parallel lines (for j = 1, 2, . . . ,M), each with a slope roughly proportional to $\Lambda_m$.

The natural logarithm of the divergence of the nearest neighbour to the j<sup>th</sup> point in the phase space is presented as a function of time. The largest Lyapunov exponent is then calculated as the slope of the 'average' line, defined via a least squares fit to the 'average' line defined by:

$$y(t) = \frac{1}{\Delta t} \langle \ln d_j(t) \rangle \tag{6}$$

where $\langle ... \rangle$ denotes the average over all values of j. This process of averaging is the key to calculating accurate values of $\Lambda_m$ using smaller and noisy data sets compared to other Lyapunov algorithms. The largest Lyapunov exponent (LLE) is 0.7586 for the HR signal shown in Figure 1.

### 3.3. The $\Lambda_m$ calculations

*The $\Lambda_m$* exponents were calculated using the Wolf and Rosenstein algorithms implemented as previously recommended. For both algorithms, the first two steps were similar. An embedded point in the attractor was randomly selected, which was a delay vector with $d_E$ elements

$$[x(t), x(t+\tau), x(t+2\tau), ..., x(t+(d_E-1)\tau)] \tag{7}$$

This vector generates the reference trajectory. It's nearest neighbour vector

$$[x(t), x(t_0+\tau), x(t_0+2\tau), ..., x(t_0+(d_E-1)\tau)] \tag{8}$$

was then selected on another trajectory by searching for the point that minimizes the distance to the particular reference point. For the Rosenstein algorithm, it is imposed the additional constraint that the nearest neighbour has a temporal separation greater than the mean period of the time series defined as the reciprocal of the mean frequency of the power spectrum.

The two procedures then differed. For the Wolf algorithm, the divergence between the two vectors was computed and as the evolution time was higher than three sample intervals, a new neighbour vector was considered. This replacement restricted the use of trajectories that shrunk through a folding region of the attractor. The new vector was selected to minimize the length and angular separation with the evolved vector on the reference trajectory. This procedure was repeated until the reference trajectory has gone over the entire data sample and $k_1$ was estimated as:

$$\Lambda_m = \frac{1}{t_M - t_0} \sum_{k=1}^{M} \ln \frac{L'(t_k)}{L(t_{k-1})} \tag{9}$$

where $L'(t_k)$ and $L(t_{k-1})$ are the distance between the vectors at the beginning and end of a replacement step, respectively, and M is the total number of replacement steps. Note this equation uses the natural logarithm function and not the binary logarithm function as presented by Wolf [7]. This change makes $\Lambda_m$ exponents more comparable between the two algorithms.

For the Rosenstein algorithm, the divergence d (t) between the two vectors was computed at each time step over the data sample. Considering that $N - (d_E - 1)\tau$ embedded points (delay vectors) composed the attractor, the above procedure was repeated for all of them and $\Lambda_m$ were then estimated from the slope of linear fit to the curve defined by:

$$y(t) = \frac{1}{\Delta t} \langle \ln d_j(t) \rangle \tag{10}$$

where $\langle \ln d_j(t) \rangle$ represents the mean logarithmic divergence for all pairs of nearest neighbours over time. . This process of averaging is the key to calculating accurate values of $\Lambda_m$ using smaller and noisy data sets compared to Wolf algorithms.

### 3.4. The modified Mazhar-Eslam Lyapunov Exponent

The modified algorithm steps are same Rosenstein's algorithm steps. However, the Rosenstein's algorithm uses the Fast Fourier Transform (FFT) as shown in figure 3; the modified algorithm uses the Discrete Wavelet Transform (DWT) as shown in figure 4. That it is because DWT advantages comparing with FFT. In the next some reasons to choose DWT instead of FFT.

#### 3.4.1. Preference for choosing DWT instead of FFT

Although the FFT has been studied extensively, there are still some desired properties that are not provided by FFT. This section discusses some points are lead to choose DWT instead of FFT. The first point is hardness of FFT algorithm pruning. When the number of input points or output points are small comparing to the length of the DWT, a special technique called pruning is often used [12]. However, it is often required that those non-zero input data are grouped together. FFT pruning algorithms does not work well when the few non-zero inputs are randomly located. In other words, sparse signal does not give rise to faster algorithm.

The other disadvantages of FFT are its speed and accuracy. All parts of FFT structure are one unit and they are in an equal importance. Thus, it is hard to decide which part of the FFT structure to omit when error occurring and the speed is crucial. In other words, the FFT is a single speed and single accuracy algorithm.

The other reason for not selecting FFT is that there is no built-in noise reduction capacity. Therefore, it is not useful to be used. According to the previous ,the DWT is better than FFT especially in Lyapunov exponent calculations when be used in HRV, because each small variant in HRV indicates the important data and information. Thus, all variants in HRV should be calculated.

#### 3.4.2. Modified calculation of Largest Lyapunov Exponent Λ

The modified method depends on Rosenstein algorithm's strategy with replacing the FFT by DWT to estimate lag and mean period. However, the modified method use the same technique of Wolf technique for $\Lambda_m$ calculating except the first two steps and the final step as they are taken from Rosenstein's method. The Lyapunov exponent (Λ) measures the degree of separation between infinitesimally close trajectories in phase space. As discussed before, the Lyapunov exponent allows determining additional invariants. The modified method of LLE (Λ) is calculated as

$$\Lambda = \sum_{i=1}^{j} \frac{\Lambda_i}{j} \quad (11)$$

Note that the $\Lambda_i$s contain the maximum Λ and variants Λs that indicate to the helpful and important data. Therefore, the modified Lyapunov is a more sensitive prediction tool. Thus, it is robust predictor for real time, in addition to its sensitivity for all time whatever the period. It is found that the modified largest Lyapunov exponent ($\Lambda_m$) is 0.4986 for the HR signal shown in Figure 1. Thus, it is more accurate than Wolf and Rosenstein Lyapunov exponent.

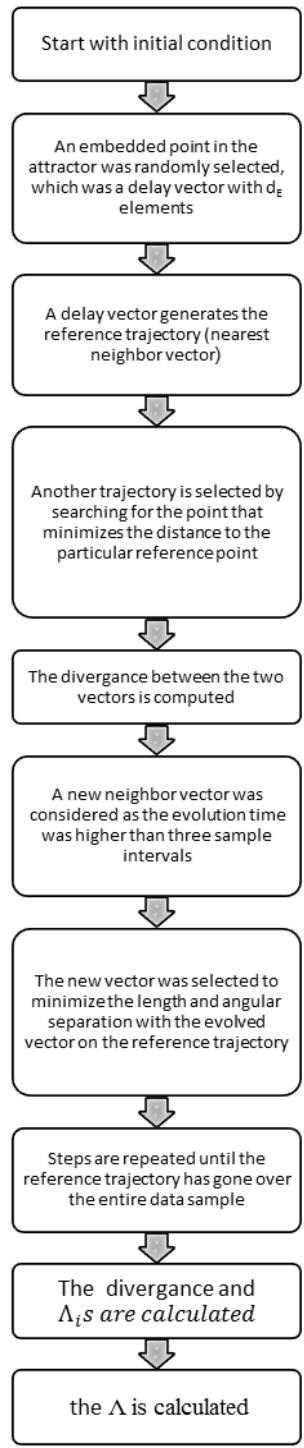

Figure 4. The flowchart of the modified algorithm.

The figure 4 shows the steps for calculating modified Mazhar-Eslam Lyapunov exponent. The modified method is the most useful and sensitive comparing to Wolf and Rosenstein methods. The table (1) discusses the different results in normal case between Modified, Wolf, and Rosenstein methods. The Rosenstein method is the lowest sensitive method because of its quite high error comparing to the optimum. The Wolf method takes a computational place of sensitive. However, the modified method shows more sensitivity than Wolf method as the modified error is lower than Wolf as shown in figure 5. The error for each case is calculated as

$$error\ (r) = |normal\ (optimum) - case| \qquad (12)$$

Thus, the accuracy for Wolf and Modified method should be calculated. The accuracy is calculated as

$$accurcy = (1 - r) \times 100\% \qquad (13)$$

Figure 6 shows the accuracy of Wolf and modified method for control or normal case. It is clear that the modified Mazhar-Eslam method is more accurate than Wolf by 0.36%. This result comes because the modified takes all Λs unlike the Wolf method as it takes only the largest. Each interval of HRV needs to be monitored and taken into account because the variant in HRV indicates to another case.

Table 1. Lyapunov result for normal case shown in figure 1

| **Optimum** | **Modified (Mazhar-Eslam) method** | **Wolf method** | **Rosenstein method** |
|---|---|---|---|
| 0.5 | 0.4986 | 0.505 | 0.7586 |
| **Error** | 0.0014 | 0.005 | 0.2586 |

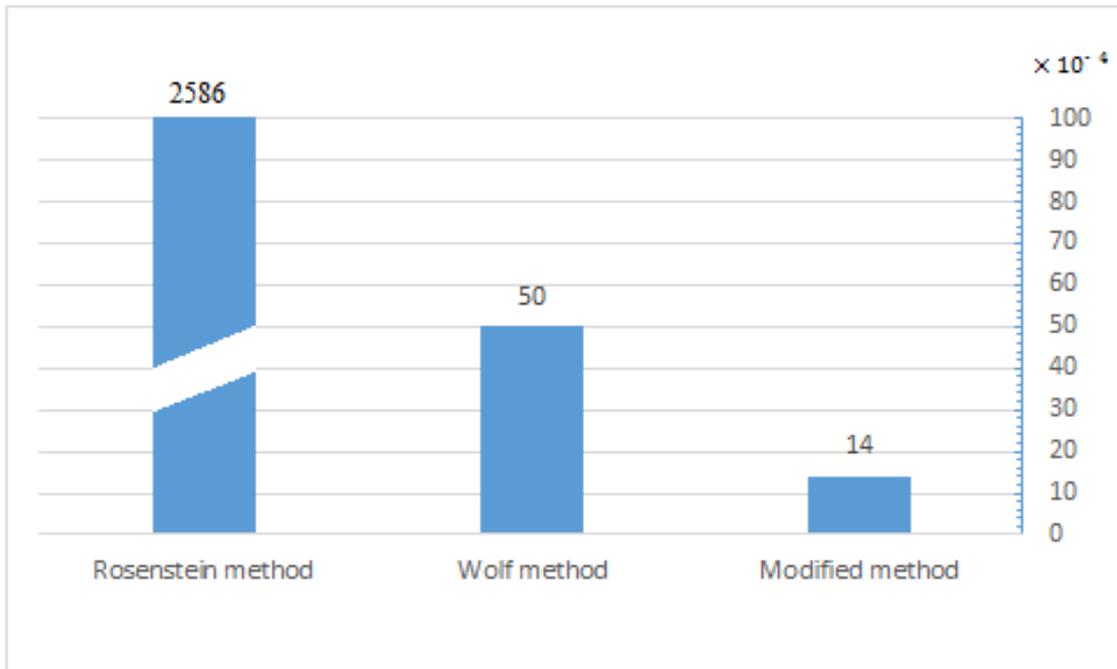

Figure 5: Methods error for the normal case.

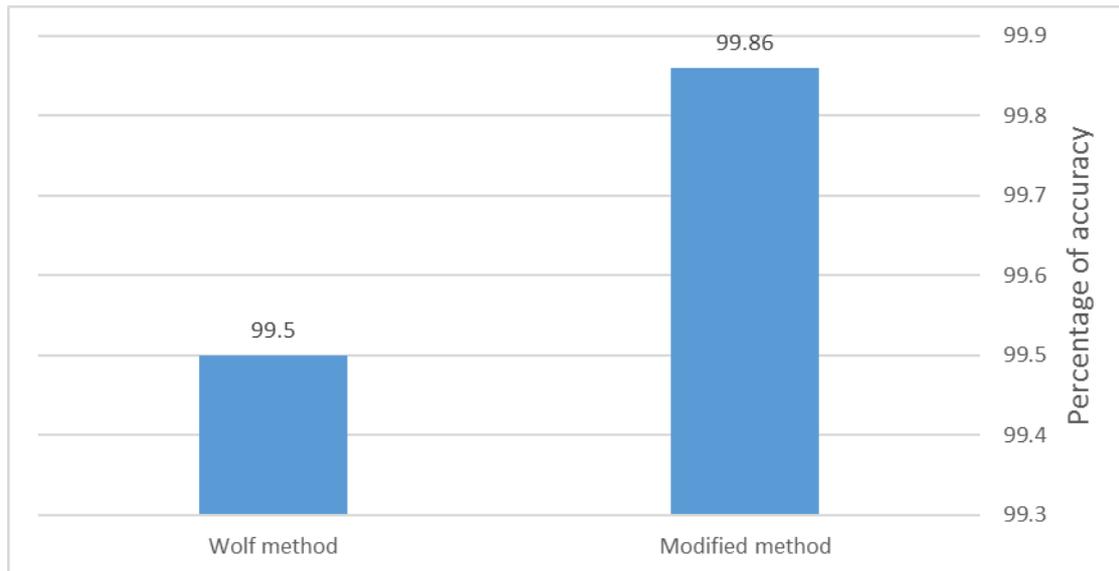

Figure 6: Accuracy percentage of Wolf and Modified method for the normal case.

## 4. COMPARISON AND VERIFYING DIFFERENT DISEASES FOR THREE METHODS OF LYAPUNOV

The data were selected; contain 9 healthy individuals males and 6 cardiac patients (two females). The data for Cardiac patients were outpatients of the Cardiology Clinic of Ege University Medical School in Turkey. The nine healthy cases consist of eight males and one female. The ages of health male cases were; age 55 years for first case, 22 years old for second and third case, 23 years old for fourth, seventh, and eighth case, 31 years for fifth case, and 21 years old for sixth case. The ninth case is female with 32 years old. The two female patients were assigned as P1 and P2. They are in the same age of 69 years old and their weight 53 kg for P1 and 40 kg for P2. The P1 had arrhythmia (Arr) plus chronic obstructive lung disease (Cold) and rheumatoid arthritis (RA). The P2 had congestive heart failure (CHF) plus cold. The others patients were males and were assigned from P3 to P6. The patient P3 is 59 years old and his weight is 70 kg. The P3 with angina pectoris (AP) had coronary by-pass surgery seven years ago and patient. Patients P4, P5 and P6 had congestive heart failure (CHF).

Approximately 4000 R-R intervals were measured from each data set. These data were recoded form ECG for more slightly than one hour on the average. The three Lyapunov methods Mazhar-Eslam, Rosenstein, and Wolf were used to predict diseases from the data set. The next table shows the different fifteen cases were predicted by different Lyapunov methods.

Table 2. Lyapunov methods result for different fifteen cases as H indicate to health case and P for patient case

| Cases | Mazhar-Eslam method | Wolf method | Rosenstein method |
|---|---|---|---|
| H1 | 0.4901 | 0.4715 | 0.7762 |
| H2 | 0.4981 | 0.4986 | 0.7865 |
| H3 | 0.5009 | 0.6751 | 0.6895 |
| H4 | 0.4999 | 0.5301 | 0.6758 |
| H5 | 0.4976 | 0.5296 | 0.8896 |
| H6 | 0.5011 | 0.5337 | 0.7836 |
| H7 | 0.5201 | 0.8439 | 0.7993 |
| H8 | 0.5126 | 0.7808 | 0.7541 |
| H9 | 0.5293 | 0.9772 | 0.9781 |
| P1 | 0.3209 | 0.4296 | 0.6675 |
| P2 | 0.2635 | 0.3687 | 0.2566 |
| P3 | 0.2501 | 0.3094 | 0.4563 |
| P4 | 0.2598 | 0.4308 | 0.6054 |
| P5 | 0.2513 | 0.2733 | 0.3325 |
| P6 | 0.2532 | 0.4365 | 0.2216 |

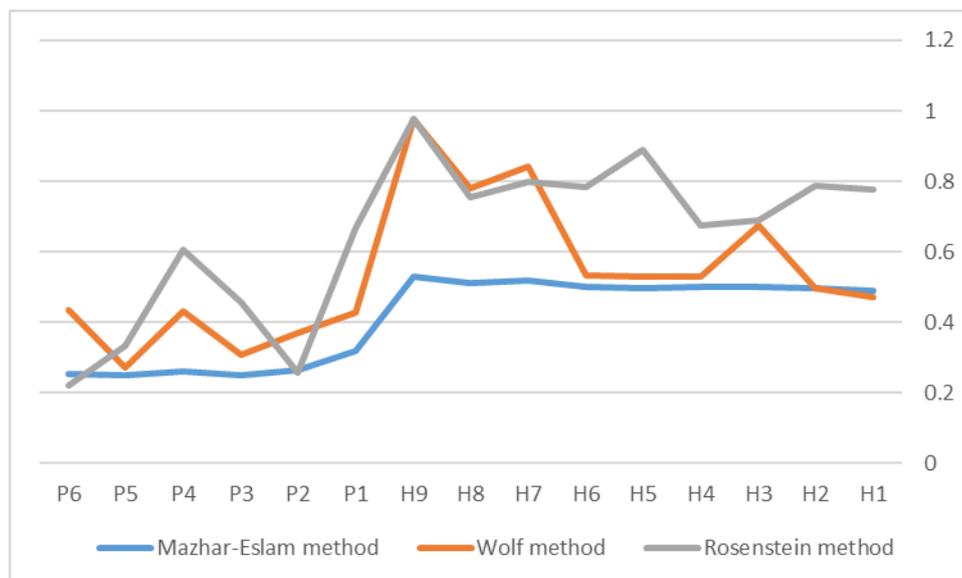

Figure 7: Stability of classification of Lyapunov methods

The figure 7 shows the comparison of three methods depends on the results in table 2. It shows that the three methods can classify the healthy and patient cases. However, for diagnosing and grouping the disease they are different. For the Rosenstein method, it has some burdens for grouped as shown. In addition, the Wolf method has some problems of grouped diseases although its sensitivity for classification. At the Mazhar-Eslam method, it shows more stability than others for classification and diagnose in HRV as shown. It is seems as a linear for the same status or condition like which appear in healthy cases and CHF patient cases.

According to previous results shown in table 2. It is clear that the Lyapunov is a sensitive tool for HRV. During observation the three methods for Lyapunov results of the cases used, they were observed that patients have reduction in HRV when compared to healthy cases.

In Rosenstein method, it is so hard to predict the disease due to some overlap results in range of patient and healthy cases. The third and fourth healthy cases were 0.6895 and 0.6758 respectively. For the first patient was 0.6675. Thus, the results were so close. Nevertheless, it is clear that healthy case in Rosenstein take the value more than 0.67 and its average result for healthy is 0.7925 as shown in figure 8. Thus, the most patient cases achieve in blew range of 0.67. The average range of CHF is 0.3865 as shown in figure 9. Also, the Rosenstein alert to the serious case when the result be lower than 0.4. Unfortunately, it is clear the difficulty for diagnosing and predicting disease although the classification for tow boundaries (healthy and patient cases). For the previous, diagnose burdens and because of utmost importance of HRV prediction and diagnosing, the Rosenstein method not recommended for critical cases in HRV although its sensitivity.

For more sensitivity in diagnosing and predicting, the Wolf Lyapunov method is used. The wolf classified the healthy cases by using the 0.45 as a threshold boundary. The Wolf Lyapunov results for healthy cases situated above of 0.45, and the patient cases were in lower than this value. Nevertheless, its average value for healthy results shown in table 2 is 0.6489 as shown in figure 8. For critical and serious cases, the Wolf values declining because the weakness and low peaks values in HRV. The mainly values of Wolf for CHF place around value of 0.43 and its average for CHF is 0.3802 as shown in figure 9. However, this method is sensitively classify, it is cannot grouped serious and critical cases. Thus, HRV needs more accurate and efficient tool for diagnosing and predicting.

The Mazhar-Eslam Lyapunov method is rebound Lyapunov exponent as sensitive tool for prediction and diagnosing. It overcomes the drawbacks in Wolf and Rosenstein methods. The table 2 shows the accuracy and efficiency of Mazhar-Eslam at many different cases. Mazhar-Eslam shows a new chapter for sensitivity and classification for HRV even these health or patient cases. For healthy cases, the evaluation criteria of statics analysis about 0.5, and its average value for result shown in table 2 is 0.5055 as shown in figure 8. In other side, the value of CHF around 0.25 and its average value depends on table 2 is 0.2548 as shown in figure 9. For the Arr the value is located in the range of 0.3. Thus, it is easy to classify and diagnose any disseises. Mazhar-Eslam method helps to predict and diagnose in real time or recorded one. It shows that it is the best method for Lyapunov exponent tool in HRV due to use all values of $\Lambda$s. The small $\Lambda$s contain many important data. Thus, Mazhar-Eslam method takes account and cares of these $\Lambda$s unlike Rosenstein and Wolf methods as they take only the maximum or largest one. The benefits of the Mazhar-Eslam method appear in serious and critical cases. It can easily predict, classify and diagnose the HRV. Therefore, the Mazhar-Eslam is best Lyapunov method for predicting and diagnosing HRV.

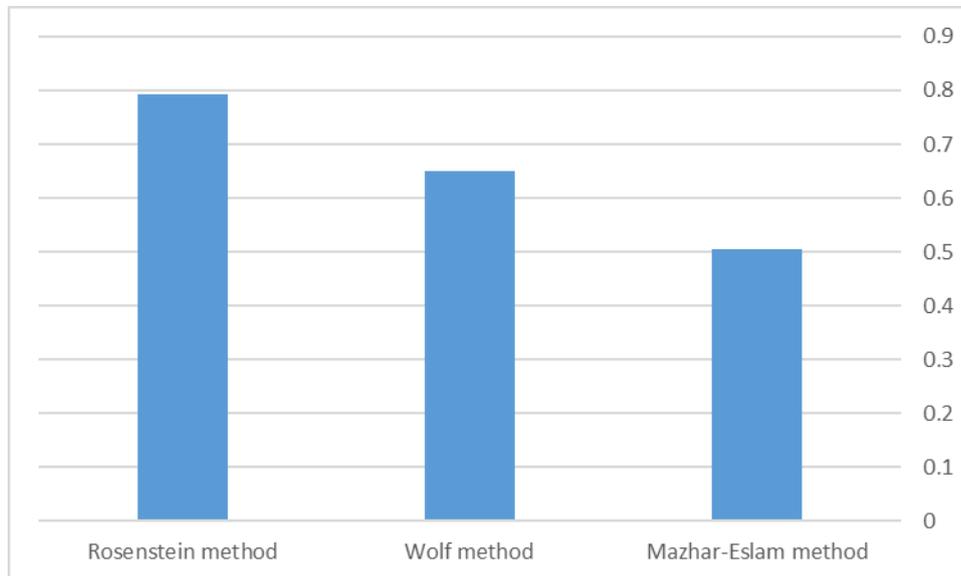

Figure 8: Average healthy results were evaluated from table 2.

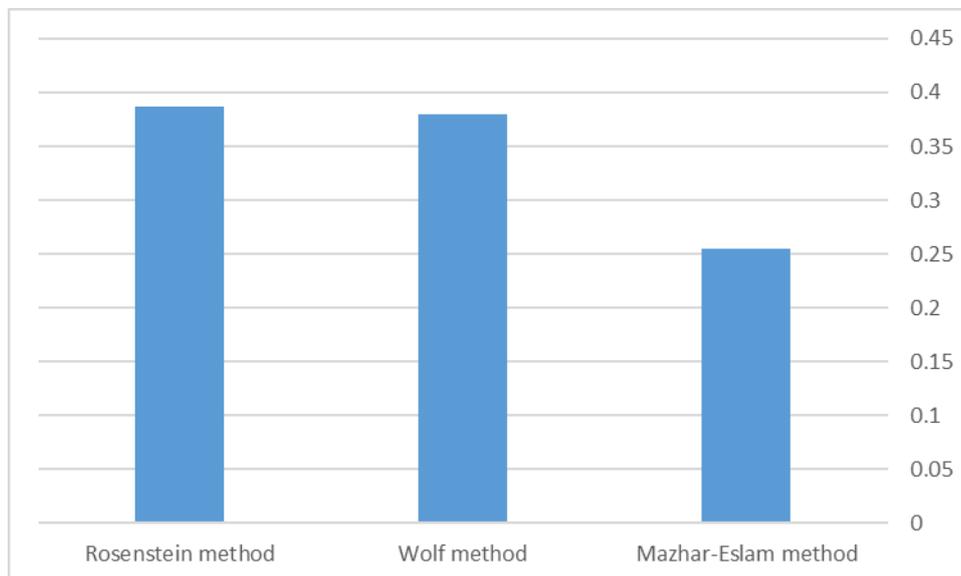

Figure 9: Average result for congestive heart failure CHF were evaluated from table 2.

Because of the heart is so sensitive to body status, the tiny error take in account. The average error for the three Lyapunov methods shown in figure 10 was calculated by using equation 12. The figure 10 shows that the Mazhar-Eslam method success to achieve the lowest error for healthy cases in Lyapunov methods as its error is 0.005522. The second lowest average error is Wolf method as it is 0.148944. The worst one is Rosenstein method as it is 0.292522.

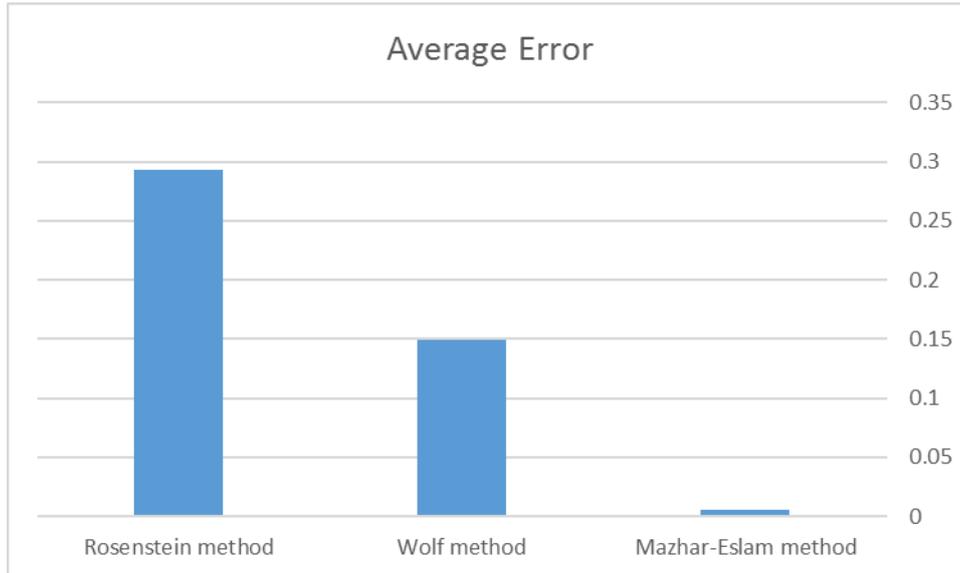

Figure 10: Average error for Lyapunov methods in healthy cases.

Thus, the accuracy for three Lyapunov methods should be calculated. The accuracy is calculated according to the equation 13. The figure 11 shows the Lyapunov methods accuracy in healthy cases. The most accurate method is Mazhar-Eslam method as it record 99.45% and the Wolf method is 85.11%. The Rosenstein method achieve 70.75 % accuracy for healthy case.

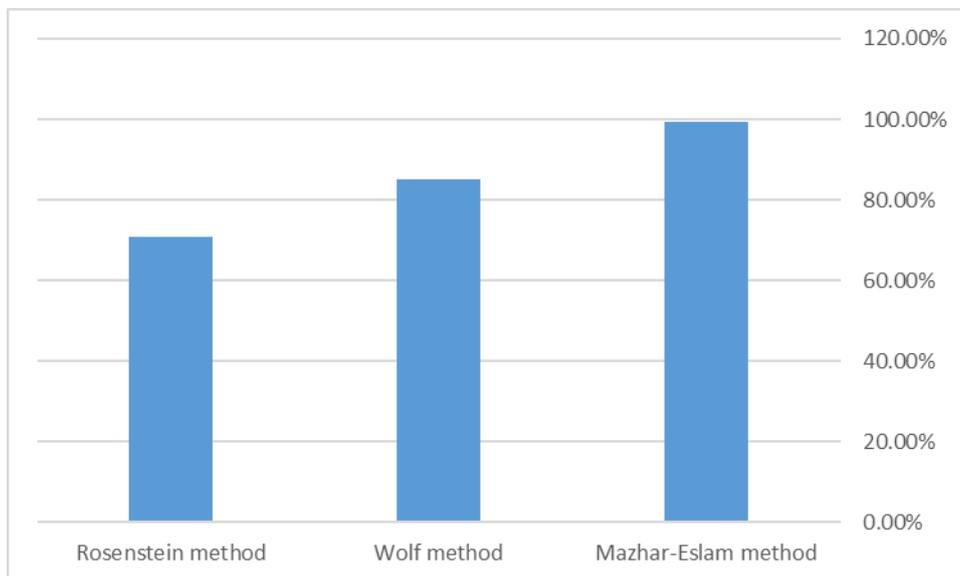

Figure 11: Lyapunov methods accuracy for healthy cases.

For all the above, the Mazhar-Eslam is strongly recommended to be used in prediction and diagnosing HRV for real and recoded time.

## 5. CONCLUSIONS

Heart Rate Variability (HRV) is reported in several cardiological and non-cardiological diseases. Also, it has a prognostic value and is therefore very important in modelling the cardiac risk. HRV is chaotic or stochastic remains highly controversial. In order to have utmost importance, HRV needs a sensitive tool to analyze it like Lyapunov exponent as it is a quantitative measure of sensitivity. While the $\Lambda_m$ from both Lyapunov exponent algorithms were nearly equal for small heart rate (HR) data sets. The Rosenstein algorithm provided less sensitive $\Lambda_m$ estimates than the Wolf algorithm to capture differences in local dynamic stability from small gait data sets. The data supported the idea that this latter outcome results from the ability and inability of the Wolf algorithm and Rosenstein algorithm, respectively, to estimate adequately $\Lambda_m$ of attractors with an important rate of convergence as those in gait. Indeed, it was found that the Wolf algorithm makes an excellent use of the attractor divergences for estimating $\Lambda_m$ while the Rosenstein algorithm overlooks the attractor expansion. Therefore, the Wolf algorithm appears to be more appropriate than the Rosenstein algorithm to evaluate local dynamic stability from small gait data sets like HRV. Increase in the size of data set has been shown to make the results of the Rosenstein algorithm more suitable, although other means as increasing the sample size might have a similar effect. The modified Mazhar-Esalm method combines Wolf and Rosenstein method. It takes the same strategy of Rosenstein method for initial step to calculate the lag and mean period, but it uses Discrete Wavelet Transform (DWT) instead of Fats Fourier Transform (FFT) unlike Rosenstein. After that, it completes steps of calculating $\Lambda$s as Wolf method. The modified Mazhar-Eslam method care of all variants especially the small ones like that are in HRV. These variants may contain many important data to diagnose diseases as RR interval has many variants. Thus, the modified Mazhar-Eslam method of Lyapunov exponent $\Lambda$ takes all of $\Lambda$s. That leads it to be robust predictor and that appear in different results between modified Mazhar-Eslam, Wolf, and Rosenstein. The Mazhar-Eslam is more accurate than Wolf and Rosenstein Lyapunov exponent. The accuracy of modified Mazhar-Eslam method for control or normal case is more accurate than Wolf by 0.36%. However, for healthy cases, the Mazhar-Eslam is more accurate than wolf by 14.34%.

## AUTHORS

**Mazhar B. Tayel** was born in Alexandria, Egypt on Nov. 20th, 1939. He was graduated from Alexandria University Faculty of Engineering Electrical and Electronics department class 1963. He published many papers and books in electronics, biomedical, and measurements.

Prof. Dr. Mazhar Bassiouni Tayel had his B.Sc. with honor degree in 1963, and then he had his Ph.D. Electro-physics degree in 1970. He had this Prof. degree of elect. and communication and Biomedical Engineering and systems in 1980. Now he is Emeritus Professor since 1999.

From 1987 to 1991 he worked as a chairman, communication engineering section, EED BAU-Lebanon and from 1991 to 1995 he worked as Chairman, Communication Engineering Section, EED Alexandria. University, Alexandria Egypt, and from 1995 to 1996 he worked as a chairman, EED, Faculty of Engineering, BAU-Lebanon, and from 1996 to 1997 he worked as the dean, Faculty of Engineering, BAU - Lebanon, and from 1999 to 2009 he worked as a senior prof., Faculty of Engineering, Alexandria. University, Alexandria Egypt, finally from 2009 to now he worked as Emeritus Professor, Faculty of Engineering, Alexandria University, Alexandria Egypt. Prof. Dr. Tayel worked as a general consultant in many companies and factories also he is Member in supreme consul of Egypt. E.Prof. Mazhar Basyouni Tayel

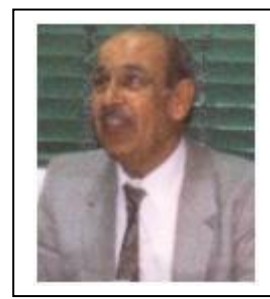

**Eslam Ibrahim ElShorbagy AlSaba** was born in Alexandria, Egypt on July. 18th, 1984. He was graduated from Arab Academy for Science Technology and Maritime Transport Faculty of Engineering and Technology Electronics and Communications Engineering department class 2007. He published many papers in signal processing.

Eslam Ibrahim AlSaba had his B.Sc. with honor degree in 2007, and then he had his MSc. Electronics and Communications Engineering degree in 2010. He is a PhD. Student in Alexandria university Faculty of Engineering Electrical department Electronics and Communications Engineering section from 2011 until now. From 2011 until now, he works as a researcher in Alexandria University. In addition, now he works as a lecturer in Al Baha International college of Science, KSA.

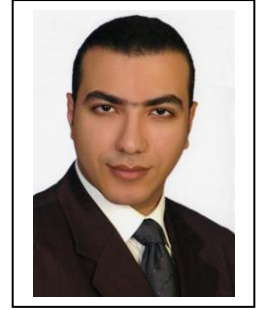